\documentclass{elsart}
\usepackage{natbib}
\usepackage{graphicx,amssymb}
\begin{document}
\begin{frontmatter}

\title{Scale-free networks with a large- to hypersmall-world transition}

\author[unm]{Petter Holme}

\address[unm]{Department of Computer Science, University of New Mexico,
  Albuquerque, NM 87131, U.S.A.}

\begin{abstract}
  Recently there has been a tremendous interest in models of networks
  with a power-law distribution of degree---so called ``scale-free
  networks.'' It has been observed that such networks, normally, have
  extremely short path-lengths, scaling logarithmically or slower with
  system size. As an exotic and counterintuitive example we propose a
  simple stochastic model capable of generating scale-free networks
  with linearly scaling distances. Furthermore, by tuning a parameter
  the model undergoes a phase transition to a regime with extremely
  short average distances, apparently slower than $\log\log N$ (which
  we call a hypersmall-world regime). We characterize the degree-degree
  correlation and clustering properties of this class of networks.
\end{abstract}

\end{frontmatter}

\section{Introduction}

A major source of the recent surge of interest in complex networks has
been the discovery that a large class of real-world networks have
distributions of degree (the number of neighbors of a vertex)
scaling like a power-law~\cite{ba:rev,doromen:book,mejn:rev}, so called
\textit{scale-free networks}. Ever since Ref.~\cite{ba:model} there
has been a tremendous number of works modeling networks with
power-law degree distributions. One characteristic feature of most such
model networks are that the distances (numbers of edges in shortest
paths between vertices) are very short (so called \textit{small-world
  networks}~\cite{mejn:rev}), scaling like a logarithm, or like an even
slower increasing function~\cite{chung_lu:pnas,cohen:ultra},
of the system size. It is however not true that all models with a
power-law degree distribution have slowly increasing distances. In
this paper we propose a simple, random network model
having a power-law degree distribution with an arbitrary exponent, and
a transition between regimes of linearly scaling distances and
distances scaling slower than a double logarithm. Our model is not as
much a model of a real-world system as an example of vast variety of
structure within the class of networks defined by a degree
distribution.

\section{The model}

The models of scale-free networks can be divided into
classes. Probably most proposed models are Markov chain
growth~\cite{ba:model,hk:model,klemm,jap:sf,kahng:sf,fkp:model,klein:web}
or equilibrium~\cite{bjk:scalefree,jensen:sf,val:soft,nong:sf} models
where the power-law degree distribution is an emergent property of the
system. In another class of models the degrees are treated as intrinsic
properties of the vertices and thus preassigned before the edges are
added.\footnote{There are also crossover models between these two
  classes---models where the degree distribution depend both on the a
  stochastic evolution process and intrinsic properties of the
  network~\cite{kahng:qc,chung_lu:pnas,biabar:bec}.} Such models include the
rather frequently used ``configuration
model''~\cite{bek:conf,molloy:cri}, a model for networks with
clear-cut core-periphery structure~\cite{my:cps} and Internet at the
largest scale~\cite{inet}. The model we propose belongs to the latter
class.

Let $V$ be the set of $N$ vertices and $E$ be the set of $M$
edges. Our algorithm is defined as follows (details will be discussed
below):
\begin{enumerate}
\item \label{step:rnd} Draw $n$ random integers (that will represent
  desired degrees) in the interval $[2,n^{1/(1-\gamma)}]$. Let the
  probability of picking $K$ be proportional to $K^{-\gamma}$.
\item \label{step:sort} Sort the integers to an increasing sequence
  $K_1\leq\cdots\leq K_n$.
\item \label{step:attach} Go though the sequence in increasing index-order and
  for each vertex $i$:
  \begin{enumerate}
  \item \label{step:attach_up} With probability $1-p$,
    go through the vertices $[i+1,n]$ in increasing order and add
    an edge $(i,j)$ ($j\in[i+1,n]$) if the degrees of $i$ and $j$,
    $k_i$ and $k_j$, are lower than $K_i$ and $K_j$ respectively.
  \item \label{step:attach_down} Otherwise (if
    step~\ref{step:attach_up} is not realized, i.e.\ with probability
    $p$) go through the vertices $[i+1,n]$ in decreasing order and add
    an edge $(i,j)$ if $k_i<K_i$ and $k_j<K_j$.
  \end{enumerate}
\item \label{step:add1} For every vertex with a degree $k_i$ less than
  its desired degree $K_i$, add $K_i-k_i$ one-degree vertices.
\end{enumerate}
The total number of vertices $N$ will be $n$ plus the number of
one-degree vertices added in step~\ref{step:add1}. By construction, the
networks will have a power-law degree distribution for degrees of two
or larger. The reason degree-one vertices are added in
step~\ref{step:add1} and not generated in step~\ref{step:rnd} is to give
all high-degree vertices their desired degree. Without
step~\ref{step:add1} there would be many vertices with $K_i>k_i$ which
would affect the high-degree vertices more than low-degree
vertices. Since most power-law network models emphasize accurate
modeling of the right end of the degree distribution, we add
degree-one vertices last. Step.~\ref{step:sort} is the computational
bottleneck in the algorithm making the execution time of network
construction $O(\max(n\log n,N))$ with a fast sorting algorithm.

\section{Numerical results}

In this section we will study the network structure numerically. We
will use $10^4$ network realizations for the averages.
For readability we use only one exponent of the
degree distribution, $\gamma=2.2$. The reason we choose this exponent
is that it is we want a small exponent to make the network as broad
(and unlike other sharper distributions) as possible; and, roughly
speaking, $2.2$ is the smallest exponent commonly seen in real world
networks (for e.g.\ the Internet~\cite{doromen:book,ba:rev}). We have
checked all results for other exponents in the interval
$2.2<\gamma\leq 3$ too, and the conclusions will qualitatively hold
for these values too. In general the networks can be composed of
disconnected components (connected subgraphs). As common in such cases
we will measure the quantities for the largest component. The scaling
of the largest component size will also be discussed.

\begin{figure}
  \center{\resizebox*{0.75\linewidth}{!}{\includegraphics{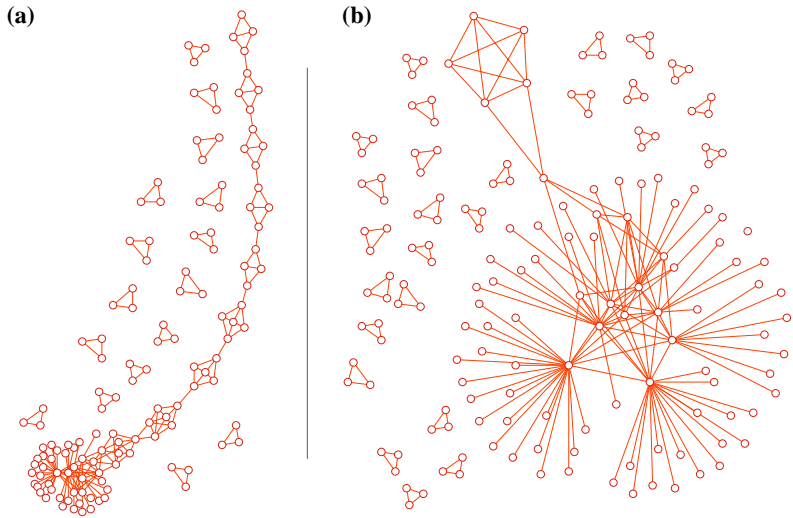}}}
  \caption{
    Example networks with $n=100$ and $p=0$ (a) and $p=0.01$ (b).
  }
  \label{fig:ex}
\end{figure}

\subsection{Example networks}

To get a feeling for the structure of the networks produced by the
model two $n=100$ networks are displayed in Fig.~\ref{fig:ex}. The
network in Fig.~\ref{fig:ex}(a), with $p=0$, is fragmented. The
largest component has a chain-like shape. This can be understood from
the construction algorithm. The major part of the degree-two vertices
will form isolated triangles. Vertex $i$ attach to $i+1$ and $i+2$,
then $i+1$ attaches to $i+2$ and all their desired degrees are
reached. Starting from degree-three vertices a component will be
formed. The degree-one vertices are connected to vertices of highest
degree. If only a small fraction of the vertices are connected in
reverse order the resulting network can be very different, see
Fig.~\ref{fig:ex}(b). Now almost the whole largest component
is directly connected to a core of a dozen, or so, vertices. Like the
$p=0$ network there is a number of disconnected degree-two vertices.
 
\begin{figure}
  \center{\resizebox*{0.75\linewidth}{!}{\includegraphics{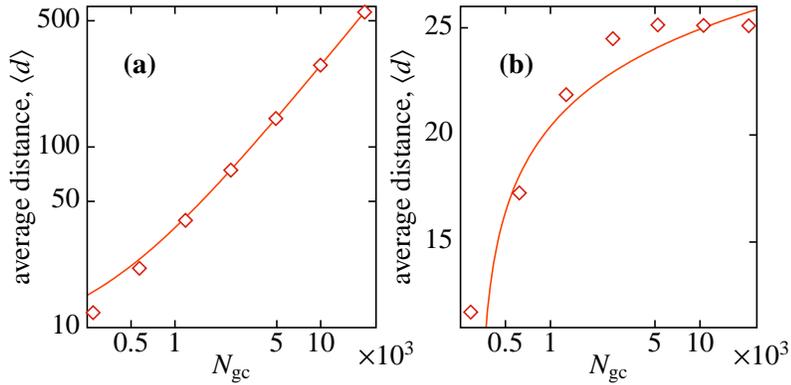}}}
  \caption{
    The length scaling. (a) shows the average distance as a function
  of the size of the largest component for $p=0$. The line is a fit to a
  linear form $8.2\pm 1.2+(0.0274\pm0.002)N$. (b) shows the
  corresponding curve for $r=0.01$. The line is a fit to a
  general $\log\log$-form: $a_1+a_2\ln(a_3+\ln(a_4+N))$ where
  $a_1,\cdots, a_4$ are constants. Note the log-log scaled axes in (a)
  and log-lin scales in (b). Errorbars are smaller than the symbol
  size.}
  \label{fig:len}
\end{figure}

\subsection{Distance scaling}

Next we turn to the central quantity for our studies---the distance
scaling. In Fig.~\ref{fig:len}(a) we plot the average distance as a
function of the size of the largest component for $p=0$. The scaling
is, to a very high accuracy, linear. Other $\gamma$-values show the
same qualitatively the scaling properties but the slope of the $d(N)$
curves is a function of $\gamma$ (lower $\gamma$ values have steeper
slopes). This is in stark contrast to other scale-free network
generators with length scaling like $\log N$, $\log N / \log\log N$ are
even $\log\log N$~\cite{cohen:ultra,chung_lu:pnas}. To
interpret this we make two trivial observations about length
scaling in network models with a fixed average degree: First, that the
average distance in the network cannot increase faster than the
maximal distance (the \textit{diameter}). Second, that the diameter
(and thus the average distance) cannot increase faster than
linearly. One can thus say that the distances in our network model
increase as fast as possible, given the average degree of the
network. If $p$ is just a tiny bit larger than zero, the scenario is
drastically different. The average distance scaling for $p=0.01$ is
plotted in Fig.~\ref{fig:len}(b). $d$ increases slower than a
logarithmic function (a logarithmic growth would appear as a line in
Fig.~\ref{fig:len}(b), due to the logarithmic $N$-axis). Indeed it
seems to grow even slower than $\log\log N$. We see this by fitting
the $d(N)$ values to a general $\log\log$-form
($a_1+a_2\ln(a_3+\ln(a_4+N))$ where $a_1,\cdots,a_4$ are
constants)---the best fit is a function that increases significantly
faster than the real curve. We also test bounded exponential and
algebraic growth forms, but neither of these fits are extremely well
to the observed curve. Indeed $d(N)$ appears to be bounded, or at least
significantly slower increasing than $\log\log N$, which
would mean our model have a transition from the theoretically maximal
(linear) to the theoretically minimal (bounded) size scaling. With
reference to the term ``ultrasmall world''~\cite{cohen:ultra} we call
sub-double-logarithmic scaling ``hypersmall world.'' This
conjecture would need further studies to be firmly
established. Deriving the functional form is non-trivial---the
probability that one reverse-adding
(step~\ref{step:attach_down} of the algorithm) should occur tends to
one (as $1-(1-p)^n$) for any non-zero $p$. The first time this occurs,
a star-graph of $O(n^{1/{1-\gamma}})$ vertices of high desired degree
will be created. But the fraction of vertices in this star-graph goes
to zero as $n^{-1/(1-1/\gamma)}$. After a few reverse-addings
vertices in the middle of the spectrum of desired degrees will be
saturated with edges and the range (in the ranking of vertices)
between the high-degree vertex and the vertex being connected will
grow rapidly. This increasing range is, phenomenologically, a possible cause
for the extremely short distance-scaling.

\begin{figure}
  \center{\resizebox*{0.75\linewidth}{!}{\includegraphics{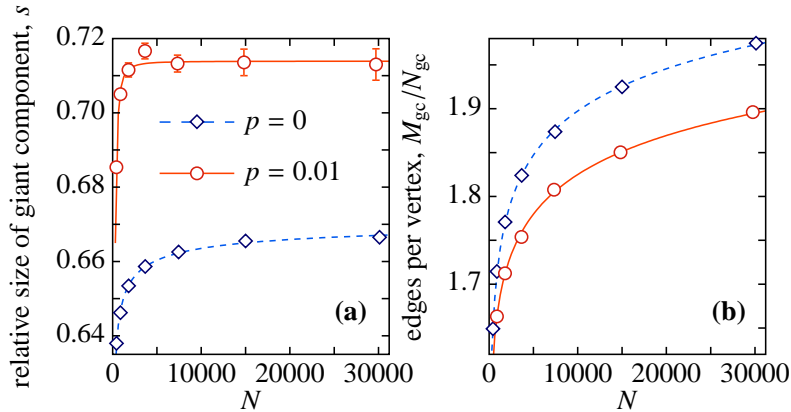}}}
  \caption{(a) shows the relative sizes of the largest
    component $s$ as a function of system size. (b) shows the density
  of edges $M_\mathrm{gc}/N_\mathrm{gc}$ (half the average degree) in
  the largest component. Both curves are fits to
  algebraic decay forms.
  }
  \label{fig:ngc}
\end{figure}

\subsection{Existence of a giant component}

As seen in Fig.~\ref{fig:ex} the model network can be
disconnected. Although our model is intended as an example of the
extreme diversity of network structure within the class of scale-free
network, we would anyway like the largest component to grow at least
linearly with the system size. In other words, there should be a giant
component in the network. In
Fig.~\ref{fig:ngc}(a) we plot the fraction of vertices in the largest
component $s$. For both $p=0$ and $p=0.01$, $s$ seems to
converge to a constant fraction of $N$. Note also that $s$ is (except
one point in the $p=0.01$ curve) an increasing function of $N$, and
it is bounded by one. Other $p$ values show the same behavior. Thus we
conclude that the model has, very likely, a giant component for all $p$
values. The density of edges in the giant component (which is half the
average degree) plotted in Fig.~\ref{fig:ngc}(b). As the size of the
giant component converges to $N_\mathrm{gc}=N^\infty_\mathrm{gc}$ the
density of edges will be bounded by $M/N^\infty_\mathrm{gc}$. The
convergence is rather slow, but consistent with an algebraic decay
form.

\begin{figure}
  \center{\resizebox*{0.75\linewidth}{!}{\includegraphics{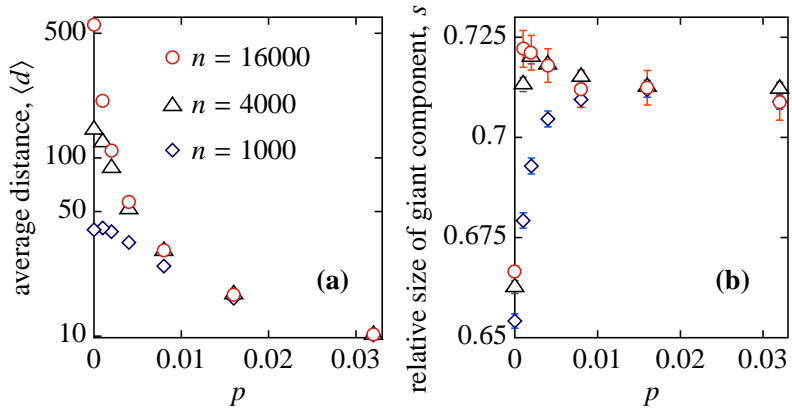}}}
  \caption{ The average distance in the largest component (a) and the
    average size of the giant component as functions of $p$.
  }
  \label{fig:vsp}
\end{figure}

\subsection{Parameter dependence of distance and largest component size}

So far, we have established that our model has two fundamentally
different network structures for $p=0$ and $p=0.01$. To complete this
picture we study the distance scaling as a function of $p$. As seen in
Fig.~\ref{fig:ngc}(a) the $\langle d\rangle(p)$-curves diverge very
slowly for other than very small $p$-values. Nothing suggests that
there would be more than two qualitatively different behaviors. Even
though $p$ represent a kind of temperature-like disorder, it is hard to say if
traditional methods of statistical physics (like renormalization
group~\cite{mejn:rg} or finite size scaling~\cite{bar:sw}) are valid
in a model with the peculiar correlations induced by our construction scheme.
To argue more speculatively, we observe that, initially,
the curves' slopes decrease increasingly fast with $p$, but at an
inflexion point the slope of a curve flattens out. This inflexion
point $\tilde{p}$ seems to
move toward $p=0$ (for $n=1000$ we have $0.004\lesssim\tilde{p}\lesssim
0.008$, for $n=4000$ we observe $0.002\lesssim \tilde{p}\lesssim 0.004$,
and for $n=16000$ we $\tilde{p}$ seem to be less than $0.001$), which
is consistent
with a $p=0$ transition. There is always a chance
$(1-p)^n$ that no reverse-adding will occur during the construction. This
is, we believe, the cause of the flattening of the $\langle
d\rangle(p)$-curves as $p\searrow 0$. (For $n=1$ and $p=0.001$---the
smallest non-zero $p$-value in Fig.~\ref{fig:ngc}(a))---absence of
reverse-adding happen in
$\sim 37\%$ of the network realizations.)

In Fig.~\ref{fig:ngc}(b) we plot the average size of the largest
component, $s$, as a function of $p$. A strongly varying $s$ would
make the distance scaling hard to interpret, but this is apparently
not the case. Instead $s$ seems to converge as $p$ increases. The
large-$p$ value is (for reasons do not speculate about in this paper)
conspicuously size-independent.

\begin{figure}
  \center{\resizebox*{0.75\linewidth}{!}{\includegraphics{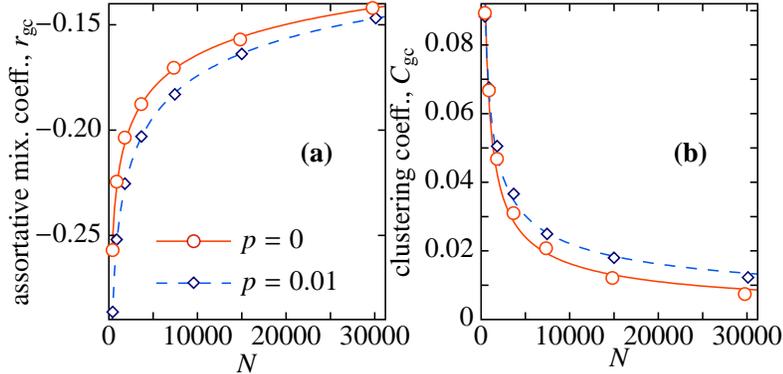}}}
  \caption{ The assortative mixing (a) and clustering (b) coefficients
    in the giant component. Standard  errors are smaller than the
    symbol size. Both curves are fits to algebraic decay forms.
  }
  \label{fig:assclu}
\end{figure}

\subsection{Degree-degree correlations}

The degree distribution is perhaps the most fundamental network
structure. A natural way to extend the characterization of the
structure of a class of networks, from the degree distribution,
 is to ask how vertices of different degrees are
interconnected. Is there a tendency for high degree vertices to attach
to other high degree vertices, or do they preferably attach to
low-degree vertices? A simple way to quantify this structure is to
measure the linear correlation coefficient of degrees at either side
of an edge, the \textit{assortative mixing
  coefficient}:~\cite{mejn:rev}
\begin{equation}\label{eq:assmix}
  r=\frac{4\langle k_1\, k_2\rangle - \langle k_1 + k_2\rangle^2}
  {2\langle k_1^2+k_2^2\rangle - \langle k_1+ k_2\rangle^2}
\end{equation}
where $k_i$ is the degree of the $i$'th argument of the edges as they
appear in an enumeration of the edges. This quantity takes values in
the interval $[-1,1]$, where low values mean that high-degree vertices
primarily attach to low-degree vertices, and high values represent a
tendency for vertices of high degree to attach to one another. In both
cases $r$ seems to converge algebraically to zero. This is seen by the
accurate fits to algebraic decay forms, $b_1N^{-b_2}$ ($b_1$ and $b_2$
are constants) in Fig.~\ref{fig:assclu} (b). This convergence to $r=0$
is more interesting when $p=0$ because, by construction, except
degree-one vertices, all vertices are attached to other vertices of
quite similar values of degree. This means there is a strong
assortative mixing in networks without degree one vertices. We
preliminary confirm that $r$ converges to a positive value if
step~\ref{step:add1} is omitted. This means the negative contribution
of attachment of degree-one vertices (at least almost) counterbalance
the positive contribution from the ordered attachment in
step~\ref{step:attach_up}. Note, though, that random networks
constrained only to a power-law degree distribution have a negative
assortative mixing coefficient~\cite{mejn:oricorr}. In other words,
in with such a null-model the degree-degree correlations are
effectively positive for our model in the $p=0$ case.

\subsection{Clustering coefficient}

Another commonly studied network structure is clustering, or the density
of triangles, in the network. In Fig.~\ref{fig:assclu} (b) we
present values of the clustering coefficient $C$---the number of
unique triangles (fully connected subgraphs of three vertices)
normalized to $[0,1]$ by dividing by the number of connected triples
of vertices (i.e.\ also including e.g.\ $\{i,i',i''\}$ where
$(i,i'),(i',i'')\in E$ but $(i'',i)\notin E$) and a combinatorial
factor three (for details see Ref.~\cite{mejn:rev}).
For both $p=0$ and $p=0.01$ the values of $C$ of the giant component
seem to converge to zero from above. Once again this can be explained
by the addition of degree-one vertices. As noted above, the highest
connected vertices are primarily connected to degree-one
vertices. Just as for the assortative mixing coefficient the
degree-one vertices plays a major role in the decreasing nature of
$C$. Without step~\ref{step:attach_up}, the clustering coefficient is
an increasing function, approaching rather large values. Now since the
difference is only the lack of degree-one vertices, the triplets
$(I,J,K)$ where $k_I=k_K=1$ and $k_J$ is very high are the main
negative contribution to $C$ (i.e., connected triples that are not
triangles).

\section{Summary and conclusions}

In this paper we have proposed a generative network model with a
power-law degree distribution. As a model parameter is
tuned the model undergoes a transition from a situation where the
average distance of the giant component scale linearly with the system
size $N$. Thus our model has a large-world regime (with
super-logarithmic distance scaling). In contrast, almost all previous
complex network models we are aware of belong to the opposite
category---small-world networks---with exponential, or
sub-exponential, distance scaling. The network models we are aware
of~\cite{wattsstrogatz,our:bipart,berker:bkt}, that do have a large-world regime,
are with one exception~\cite{klemm,bogu} trivial regular lattices or
circulants. We proceed to evaluate the structure of these model in
both the large- and small-world region of the parameter space. The
networks are disconnected even in the $N\rightarrow\infty$ limit, but
they do have a giant component.

This model is primarily intended as an exotic example, illustrating
the great variety of networks having a power-law degree
distribution. Even though the ensemble of all networks with a
power-law degree distribution never show distances growing faster than
a logarithm, this does not apply to all network models with an
emergent power-law degree distribution. Our conclusion is thus (like
in e.g.\ Ref.~\cite{hot:inet}) that one cannot treat all network
models generating power-law degree distribution as one class. Our
model include randomness in different steps of the construction, but
the sorting of vertices and sequential addition induce strong
correlations that give rise to the exotic length scaling properties.
The construction mechanism is not limited to power-law degree
distributions. We believe the high-degree vertices of power-law
networks are needed to create the hypersmall-world scaling (extremely
short pathlengths need an almost fully connected core that is directly
connected to most of the rest of the network---with too limited degrees,
such a core would be far from fully connected).
One may wonder why real-world
networks with a power-law like degree distribution (at least all we
know of) are not large-world networks. This is of course a question
for every system individually, but part of the answer can be found in
the transition found in the model. Only a tiny perturbance is needed
to turn the large-world distance scaling into a small-world---the
large-world regime is exotic, but it is also instable to small
perturbances.

\ack{
  The author acknowledges financial support from the Wenner-Gren
  foundations, the National Science Foundation (grant CCR--0331580)
  and the Santa Fe Institute; and thoughtful comments from Marian
  Bogu\~{n}a and Etsuko Nonaka.
}

\end{document}